\documentclass[prb,twocolumn]{revtex4}
\usepackage{graphicx}
\usepackage{color} 
\usepackage{transparent}
\usepackage{hyperref}

\def\cco {CaCuO$_2$}
\def\ccoc {Ca$_2$CuO$_2$Cl$_2$}
\def\cnoc {Ca$_2$NiO$_2$Cl$_2$}
\def\cueight {$2\sqrt{2}\times2\sqrt{2}\times1$}
\def\cufour {$2\times2\times1$}

\begin{document}
\title[Doped cuprates from QMC]{Ground state of doped cuprates from first principles quantum Monte Carlo calculations}
\author{Lucas K. Wagner}
\email{lkwagner@illinois.edu}
\address{Dept. of Physics, University of Illinois at Urbana-Champaign}
\begin{abstract}
The author reports on new high-fidelity simulations of charge carriers in the high-T$_c$ cuprate materials using quantum Monte Carlo techniques applied to the first principles Hamiltonian.
With this high accuracy technique, the doped ground state is found to be a spin polaron, in which charge is localized through a strong interaction with the spin.
This spin polaron has calculated properties largely similar to the phenomenology of the cuprates, and may be the object which forms the Fermi surface and charge inhomogeneity in these materials.
The spin polaron has some unique features that should be visible in X-ray, EELS, and neutron experiments.
The results contained in this paper comprise an accurate first principles derived paradigm from which to study superconductivity in the cuprates.
\end{abstract}

\maketitle

\section{Introduction}

Understanding high temperature superconductivity in the cuprates is a long-standing and major challenge in condensed matter physics.
The holes are the quasiparticles from which theories of the superconductivity are made, and in fact there have been many attempts to study doped holes using various computational and theoretical techniques.  
The key question is the nature of the holes upon doping from the antiferromagnetic insulating state.
The combination of calculations and experiments have resulted in a substantial amount of understanding of the holes in cuprates.
An important concept is spontaneous electron localization, proposed early on by authors such as Zhang and Rice\cite{zhang_effective_1988} and Emery and Reiter\cite{emery_mechanism_1988}. 
These early proposals have been followed up by other authors suggesting mechanisms for electron localization\cite{martin_electronic_1996,mott_spin-polaron_1990,lau_high-spin_2011,patterson_small_2008,hozoi_renormalization_2007}.
The literature on this subject is extensive and summarized in a number of reviews\cite{hashimoto_energy_2014,dagotto_correlated_1994,damascelli_angle-resolved_2003,kresin_colloquium:_2009,scalapino_common_2012,tsuei_pairing_2000}.

On the experimental side, angle resolved photoemission spectroscopy (ARPES)\cite{damascelli_angle-resolved_2003,razzoli_fermi_2010} and quantum oscillation\cite{doiron-leyraud_quantum_2007} measurements have elucidated the Fermi surface evolution as a function of doping, and generally agree on a Fermi surface that starts near $(\pi/2,\pi/2)$ in the crystallographic Brillouin zone and expands to a large pocket with doping.
Optical spectroscopy\cite{cilento_search_2013} has found that, in addition to a Drude peak as the system becomes metallic, additional absorption in the infrared part of the spectrum appears at around 1 eV.
Scanning tunneling microscopy(STM) has established that the system is inherently inhomogeneous\cite{fujita_simultaneous_2014,mashima_electronic_2006,wise_charge-density-wave_2008}.
Finally, neutron\cite{tranquada_evidence_1995}, x-ray\cite{abbamonte_spatially_2005}, and STM techniques\cite{fujita_simultaneous_2014} have found that these holes can arrange in stripe-like patterns.

A major theoretical challenge is that traditional electronic structure methods like density functional theory (DFT) suffer from severe errors in treatment of correlation. 
There have been studies using LDA+U\cite{rivero_general_2013,pesant_dft+u_2011} and hybrid functionals\cite{patterson_small_2008}, and using quantum chemistry techniques on cluster representations\cite{hozoi_renormalization_2007,martin_electronic_1996}. 
Recently, DFT+DMFT has been applied as well\cite{weber_strength_2010}.
The pictures emerging from these calculations have many similarities and many differences as well. 
For example, while most techniques do correctly obtain holes occupying mostly the oxygen states, they disagree on whether the interaction with the copper spins is ferromagnetic or antiferromagnetic in nature.
There is a missing element in the theoretical techniques: a variational, explicitly correlated, {\it ab-initio} calculation of the bulk material, in order to separate proposals for hole states.

In this article, I use highly accurate quantum Monte Carlo calculations of the {\em ab-initio} electronic structure of holes in the cuprates. 
This technique treats important short-range electron correlations accurately, which allows us to obtain a perspective on the effective low-energy electronic structure.
It is also variational, which allows one to test different proposals for the hole state on equal footing.
From these calculations, it appears that the most likely model for holes in cuprates is similar to one proposed by Emery and Reiter\cite{emery_mechanism_1988}. 
I will show that this hole state is consistent with the measured Fermi surface and optical spectrum, and can accomodate the formation of stripes.

The main purpose of this work is to establish the low-energy electronic structure of a hole in the cuprate material. 
At the level of accuracy currently feasible, quantum Monte Carlo techniques have the advantage that they are completely first principles and treat localized correlation very well; however, highly multiconfigurational wave functions are computationally out of reach given current capabilities. 
This work is thus concentrated on understanding the 'building blocks' of the low-energy physics of the cuprates and should serve to inform effective models of their behavior.
That is, if the ground state is highly multiconfigurational, then these single determinant-like states should be an important part of the correlated ground state.

\section{Method}

Fixed node diffusion Monte Carlo (FN-DMC) is a state-of-the-art method to calculate the electronic structure of materials from first principles, and has recently been found to have high accuracy for the undoped cuprates\cite{foyevtsova_ab_2014,wagner_effect_2014}.
Starting with a trial function $|\Psi_T\rangle$, the ground state is projected out by applying the imaginary time operator $\exp(-\hat{H}\tau)$. 
Exact projection suffers from the sign problem, which causes the method to scale exponentially in the system size.
The sign problem can be avoided by making the fixed node approximation, in which the zeros of the solution are constrained to be the same as the zeros of the trial wavefunction.
This introduces a dependence on $\Psi_T$, and the energy obtained is a variational upper bound to the true ground state energy.
In this work, many different $\Psi_T$'s are considered to estimate the best approximation to the ground state of a hole in the cuprates.
Details of the calculations are very similar to Ref~\cite{wagner_effect_2014} and are recorded in the Supplemental Information.

The first principles Hamiltonian was used:
\begin{equation}
  \hat{H}=\frac{1}{2}\sum_i \nabla_i^2 + \sum_{i<j} \frac{1}{r_{ij}} +
       \sum_{i\alpha} V_\alpha (r_{i\alpha}),
  \label{eqn:hamiltonian}
\end{equation}
$i,j$ are electron indices, $\alpha$ is a nuclear index, and the nuclear-nuclear interaction has been omitted for brevity.
Effective core potentials from Burkatski et al.\cite{burkatzki_energy-consistent_2007,burkatzki_energy-consistent_2008} were used to eliminate the core electrons and give the form for $V_\alpha$. 
Density functional theory calculations were performed using CRYSTAL\cite{dovesi_crystal:_2005} to produce a starting Slater determinant, which was allowed to break spin symmetry to form localized moments. 
The determinant was varied by changing the starting magnetic order, which resulted in determinants with different arrangements of local moments (see Figure~\ref{fig:energetics}), and by using hybrid density functional theory calculations with a varying mixing parameter.
Doped systems were simulated by removing one electron from a unit cell and compensating with a uniform background charge; the minimum cell size for a given doping percentage was used.
The QWalk\cite{wagner_qwalk:_2009} package was used to perform the quantum Monte Carlo calculations.
The Slater determinant was multiplied by a two-body Jastrow factor, which was variance optimized. Diffusion Monte Carlo was then performed using the Slater-Jastrow wavefunction as $\Psi_T$.

\begin{figure}
  \begin{tabular}{cc}
    \includegraphics{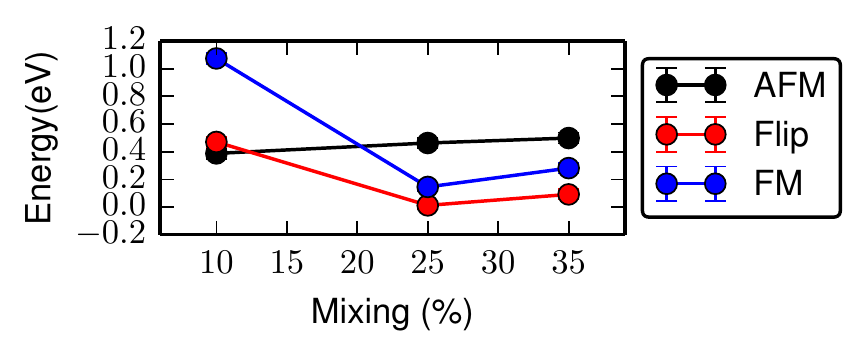}   \\
    {\bf (a) } \\
    \includegraphics{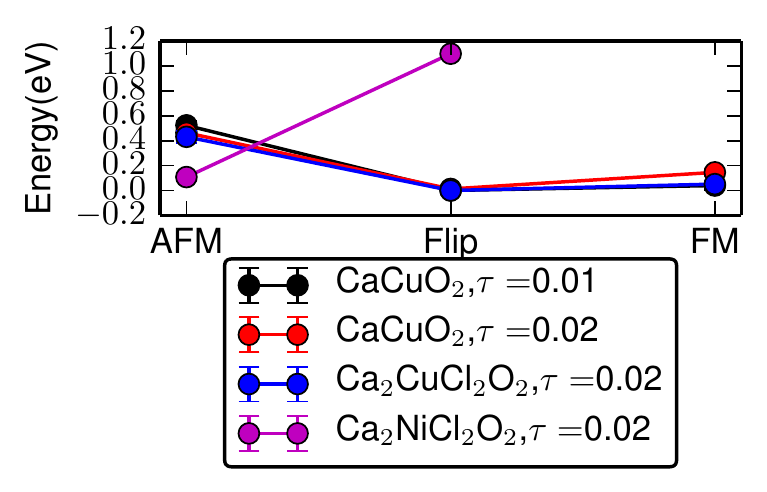} \\
    {\bf (b) } \\
  \end{tabular}
    \caption{Assessing errors in the calculations for a \cufour{} supercell, with $x=0.25$. {\bf(a)} The dependence of the relative magnetic energies versus the density functional used to generate the nodes. All structures have a minimum within stochastic uncertainties at around 25\% mixing. {\bf(b)} At 25\% mixing, the dependence on time step and on the interlayer. The physics is not qualitatively changed by these parameters. Stochastic uncertainties are approximately the size of the symbols. }
    \label{fig:errors}
\end{figure}

In diffusion Monte Carlo, there are a number of parameters that determine the accuracy.
All major parameters have been checked to the highest degree possible (Figure~\ref{fig:errors}). 
The timestep and finite size was varied, with no changes within stochastic errors.
The nodes in the input Slater determinant were varied using different hybrid density functional theories to generate the orbitals; this tuning adjusts between localized and delocalized electronic structure.
The minimum energy nodal structure was taken, which was always at 25\% mixing.
Finally, the dependence on the interlayer was checked by considering \cco{} and \ccoc{} structures, with no change in the energy differences within stochastic uncertainties.
As shown in Figure~\ref{fig:errors}, these parameters are converged.
While the solution is not exact, in particular, long-range multiconfigurational character is not captured in this technique, these calculations are the highest accuracy {\it ab-initio} results for a hole in the cuprates.
The rest of the results in this paper will be for \cco{} with $\tau=0.02$ Hartree$^{-1}$, Slater determinants generated with the PBE functional at 25\% exact exchange mixing\cite{perdew_generalized_1996,adamo_toward_1999}, twisted boundary conditions over real twists, and T-moves\cite{casula_beyond_2006}, unless otherwise indicated.

\section{Results}

\subsection{Total energy}

\begin{figure*}
  \begin{center}
  {\bf (a)}
    \raisebox{-0.5\height}{\includegraphics{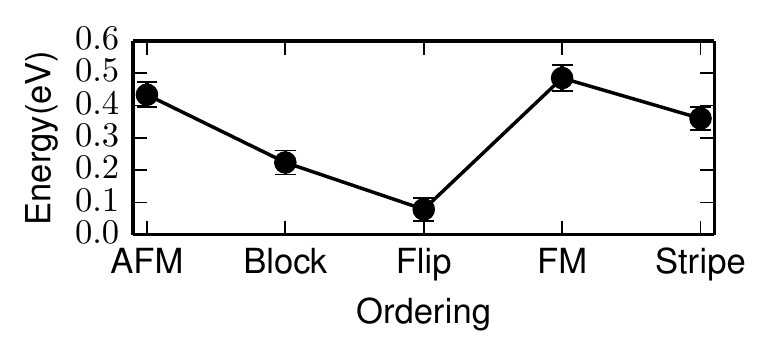}}\\
  
  \begin{tabular}{cccc}
    \multicolumn{4}{c}{Charge density {\bf(b)}}  \\
    \raisebox{-0.5\height}{\includegraphics[width=0.2\textwidth]{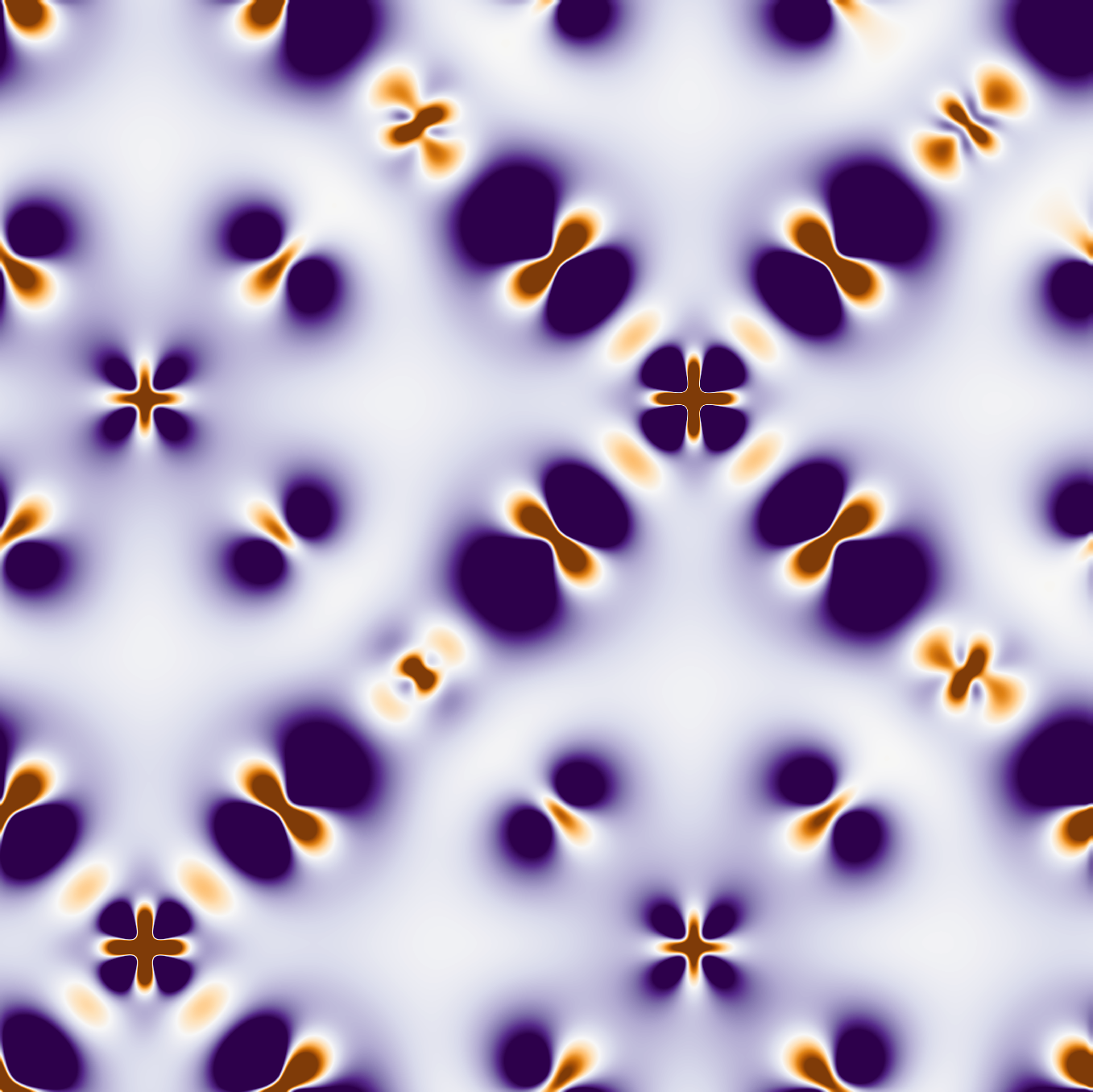}}  & 
    \raisebox{-0.5\height}{\includegraphics[width=0.2\textwidth]{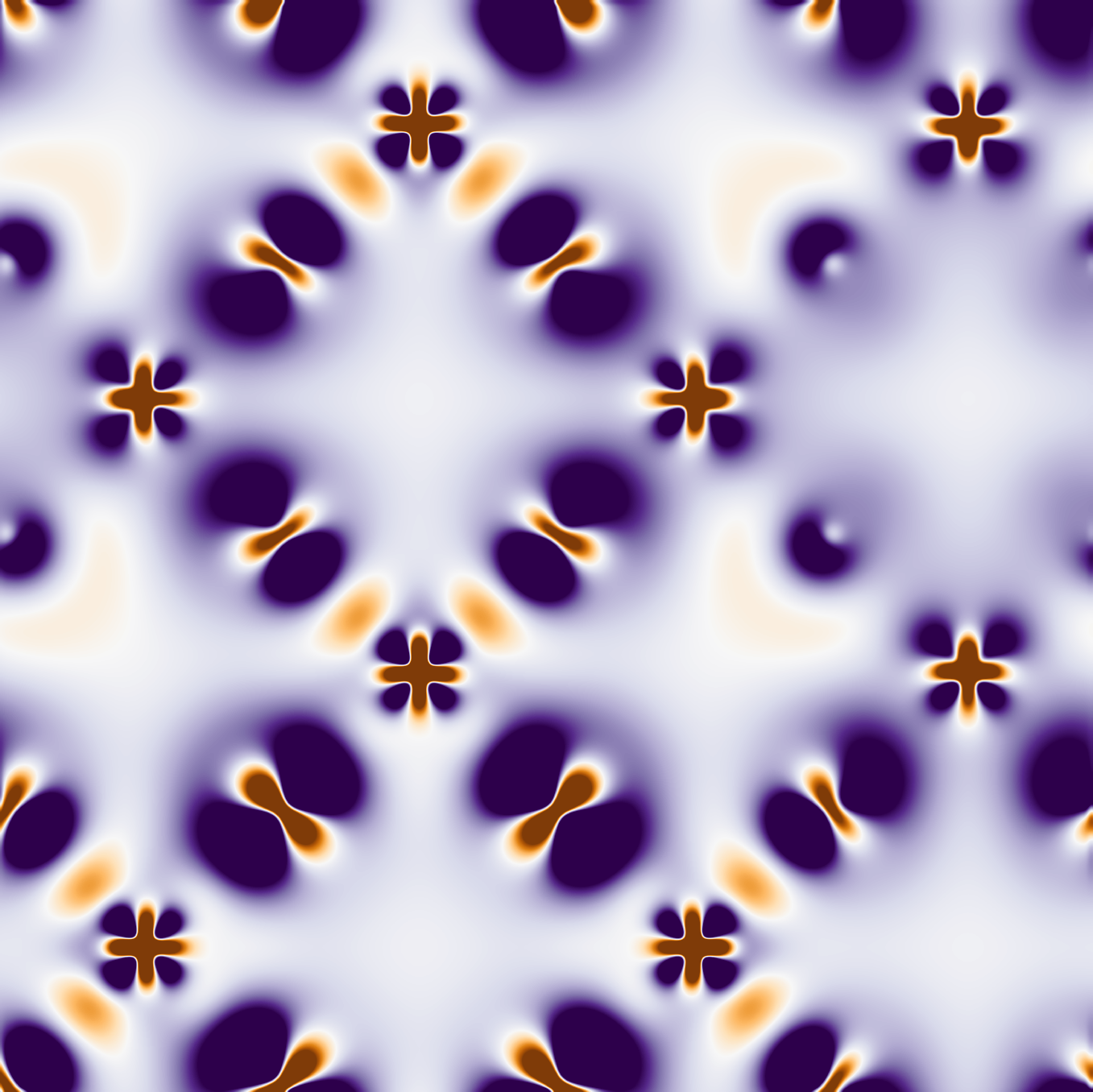}} &
    \raisebox{-0.5\height}{\includegraphics[width=0.2\textwidth]{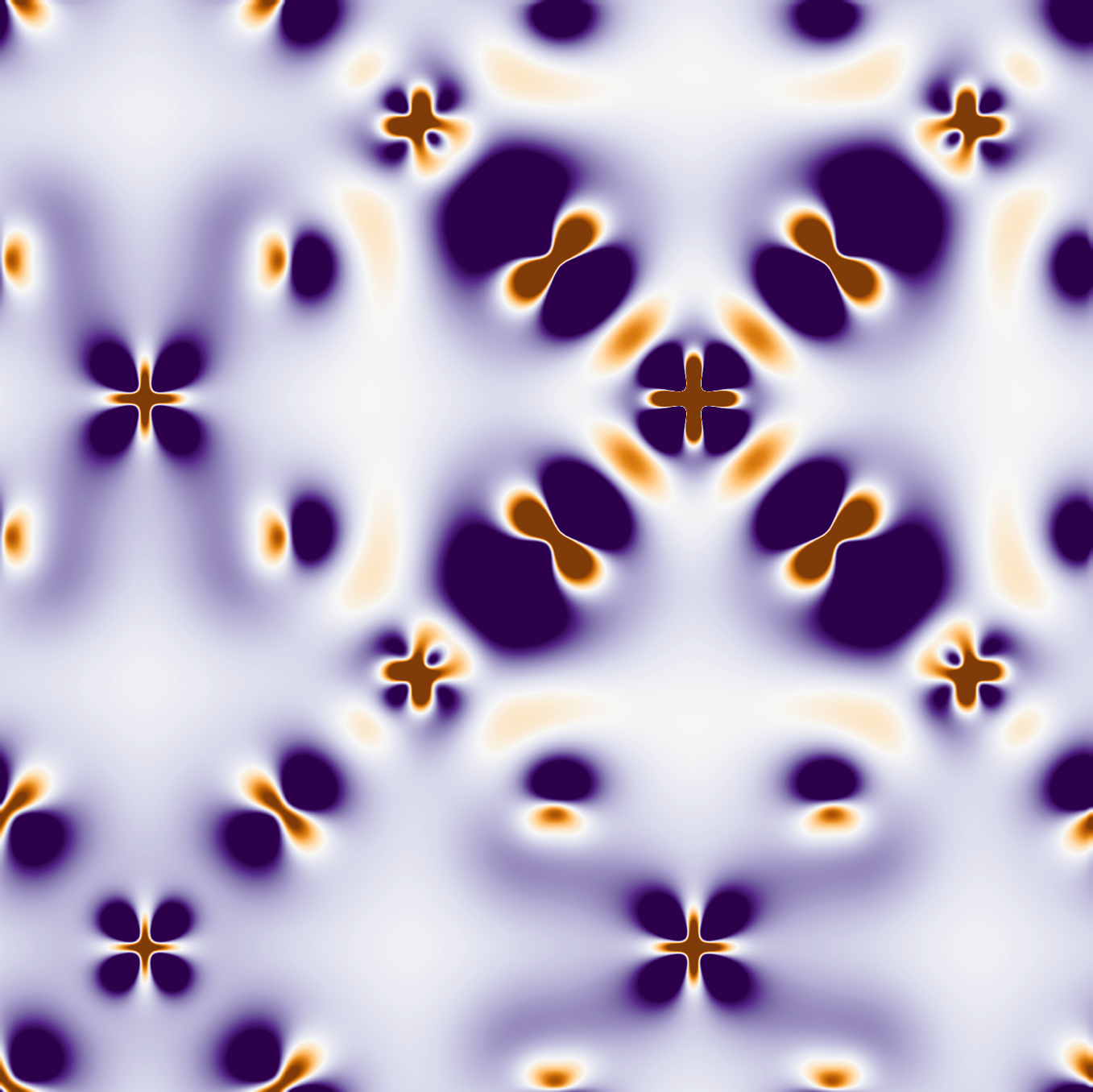}} &
    \raisebox{-0.5\height}{\includegraphics[width=0.2\textwidth]{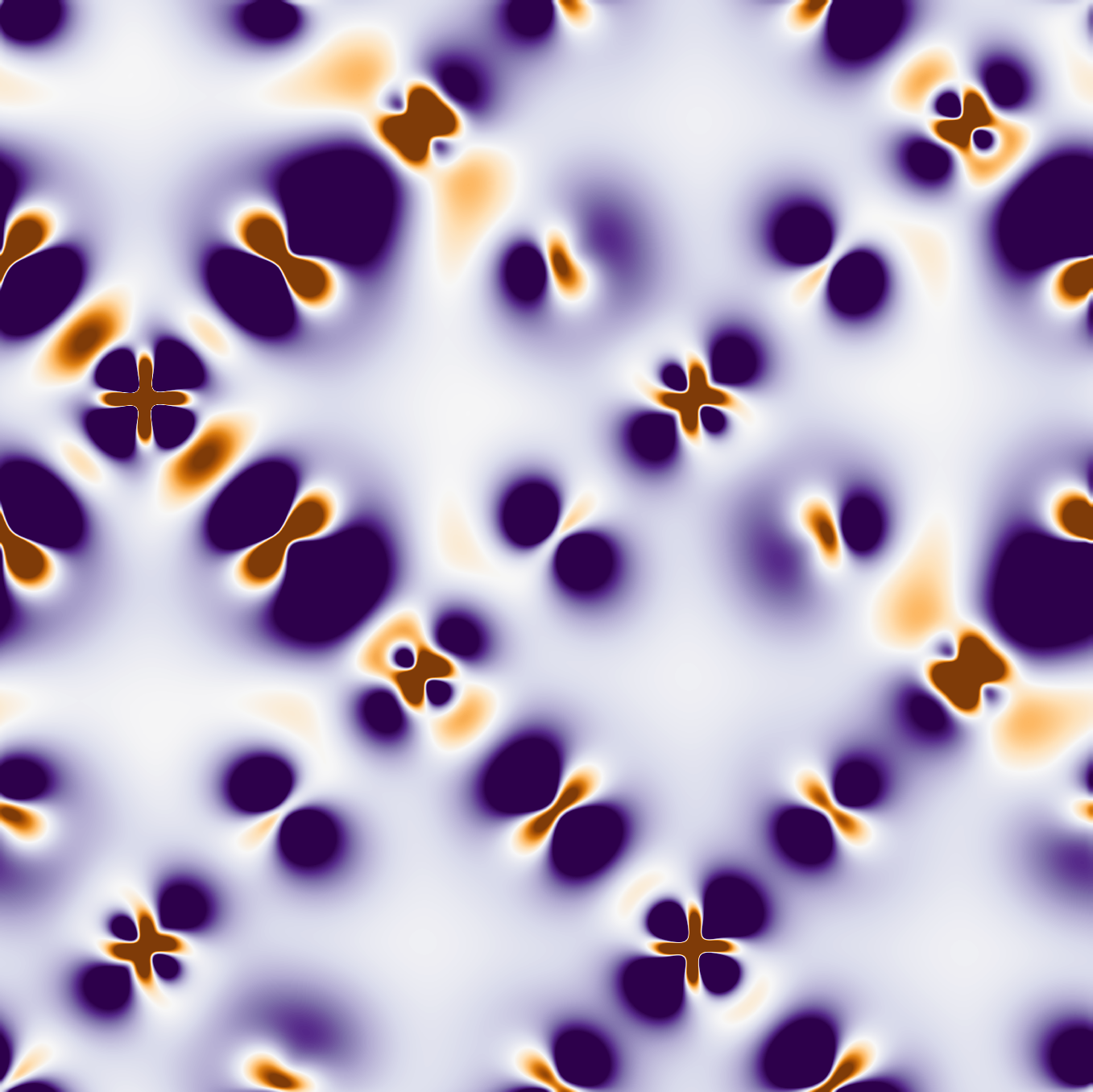}} \\
    \multicolumn{4}{c}{Spin density {\bf(c)}}  \\
    \raisebox{-0.5\height}{\includegraphics[width=0.2\textwidth]{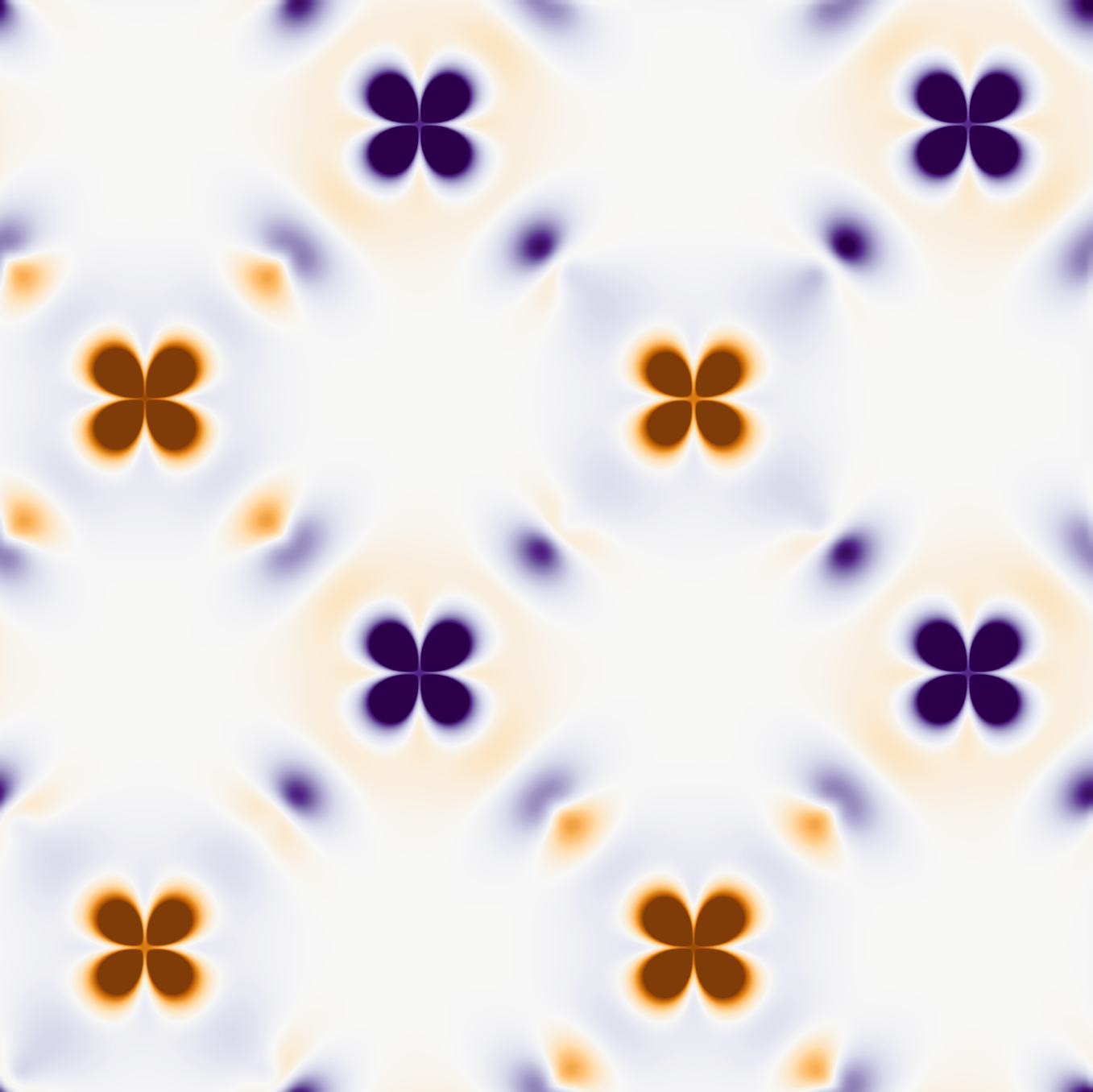}}  & 
    \raisebox{-0.5\height}{\includegraphics[width=0.2\textwidth]{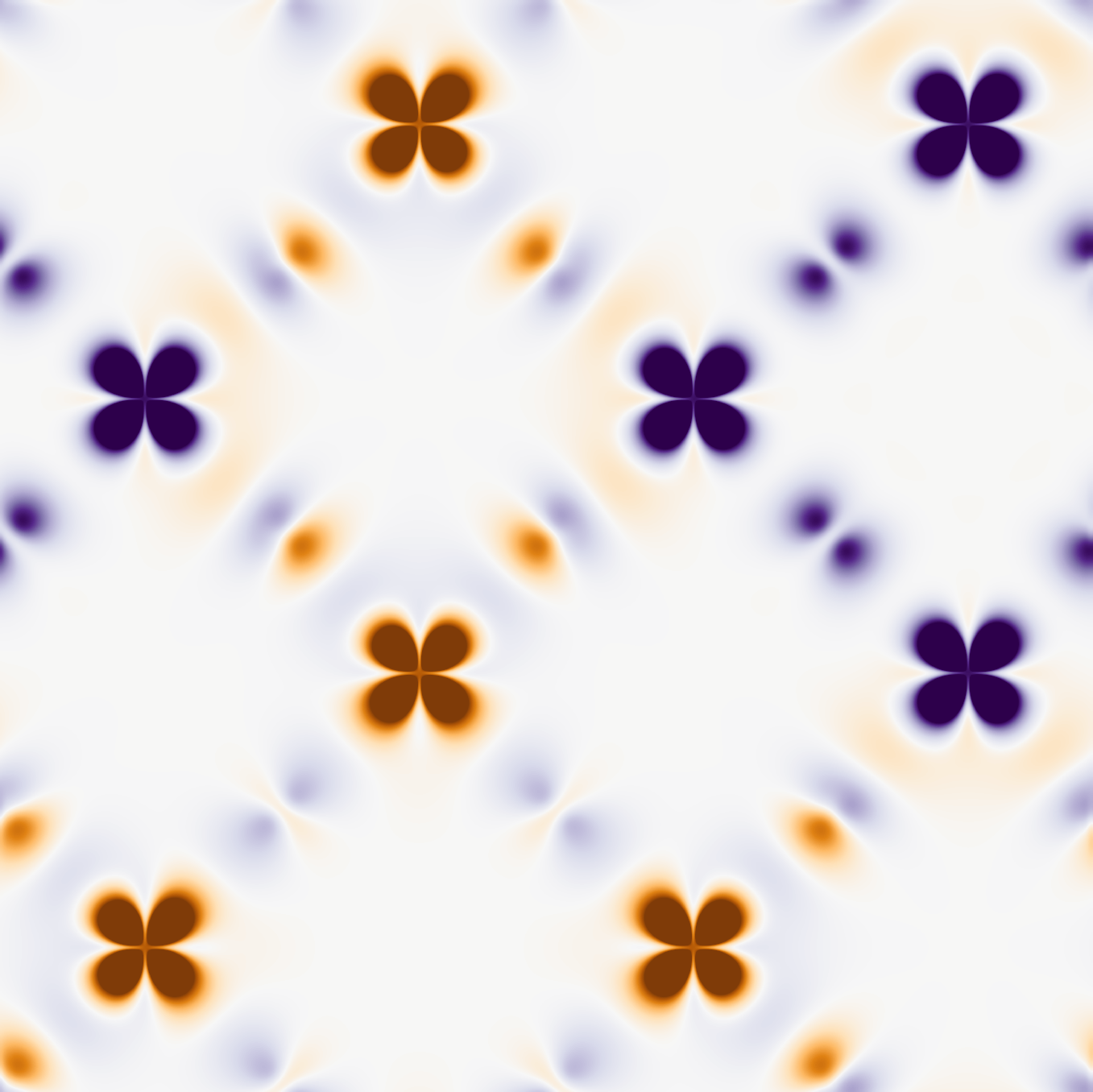}} &
    \raisebox{-0.5\height}{\includegraphics[width=0.2\textwidth]{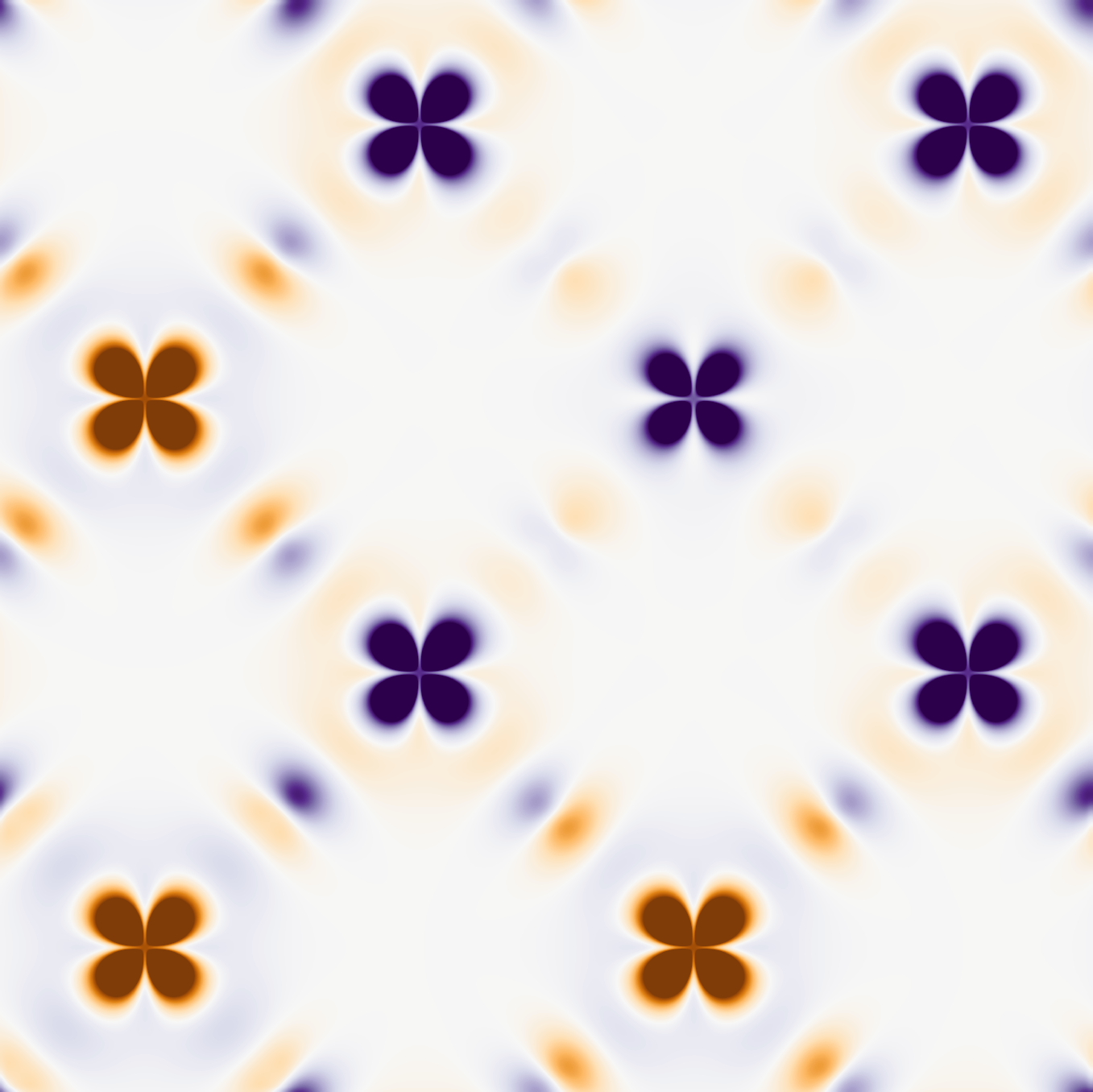}} &
    \raisebox{-0.5\height}{\includegraphics[width=0.2\textwidth]{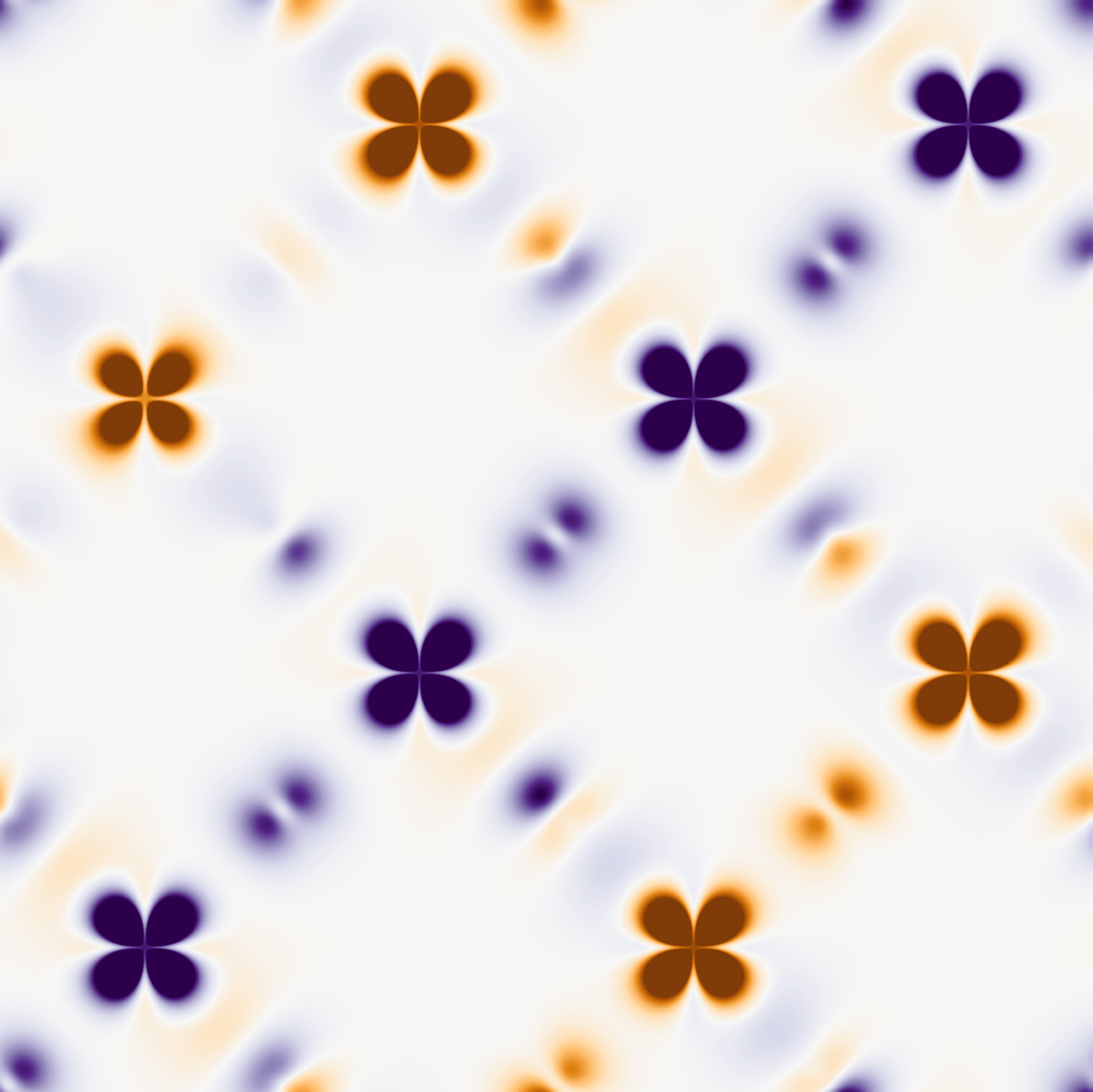}} \\
    AFM & Block & Flip & Stripe \\
  \end{tabular}
\end{center}
\caption{{\bf(a)} FN-DMC energies of different magnetic orderings considered for a \cueight{} unit cell of CaCuO$_2$.
{\bf(b)} Hole charge density obtained by subtracting the charge density of the $x=0.125$ system from the $x=0.00$ antiferromagnetic ordering. 
{\bf(c)} Spin density of the corresponding orderings for $x=0.125$. Both are calculated from the optimal single Slater determinant for that spin configuration for \cco{} and projected onto the $ab$ crystallagraphic plane. Correlations do not affect these pictures within statistical noise. In the density maps, blue is positive and red is negative. The density maps are normalized to the same value.}
  \label{fig:energetics}
\end{figure*}

Figure~\ref{fig:energetics} contains a summary of energetics for trial wave functions that differ in their magnetic ordering, for $x=0.25$ and the \cueight{} supercell.
The 'cloverleaf' shapes are the location of the Cu atoms, with oxygen atoms between them.
The hole density is calculated by subtracting the doped charge density from the undoped AFM-ordered charge density.
No matter the magnetic ordering, the hole density is largely situated on the oxygen atoms, in agreement with X-ray experiments. 
Changing the magnetic ordering affects the distribution of hole charge.

An immediately striking result in Figure~\ref{fig:energetics} is that the flipped configuration, which has a single copper atom with spin reversed from the checkerboard AFM pattern, is lowest in energy.
The flipped spin creates a region of five copper atoms with aligned spins, and a hole is attracted mostly to the oxygen atoms between the spin-aligned atoms.
The `flip' configuration is a spin polaron. 
The closely related compound, \cnoc{} does not exhibit this effect (Figure~\ref{fig:errors}), so it appears to be a special feature of the cuprates.

\subsection{Excitation properties}

\begin{figure}
  \includegraphics{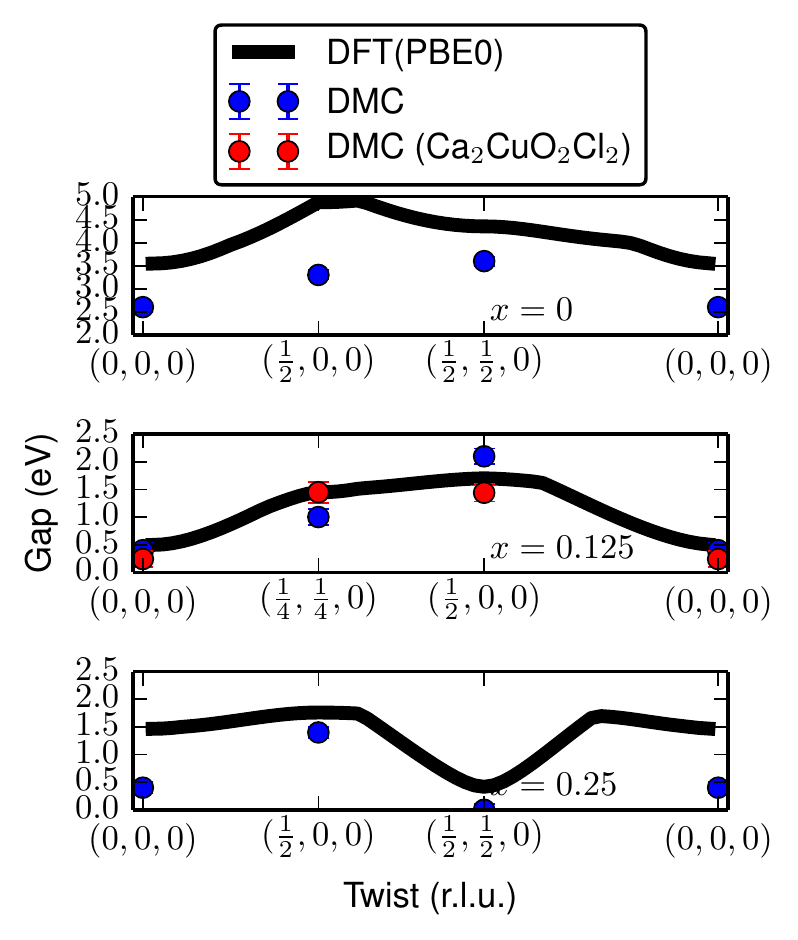}

  \caption{Optical gaps in FN-DMC and PBE0 calculated by promoting an electron in the Slater determinant, performed in the \cufour{} cell for $x=0.00$ and $x=0.25$, and in the \cueight{} cell for $x=0.125$. The twisted boundary condition of the wave function in the crystallographic primitive cell is labeled on the horizontal axis.
  }
  \label{fig:gap}
\end{figure}

The gap was computed in FN-DMC by promoting an electron in the trial Slater determinant from the highest occupied state to the lowest unoccupied state. 
In the case of \ccoc, the PBE0 ordering is incorrect, and the promotion was performed from the second-highest one-particle state.
For the doped configurations, both spin channels were attempted and the lower energy excitation was chosen. 
In Figure~\ref{fig:gap}, the gap as a function of the twisted boundary condition in the supercell are presented for the lowest energy magnetic configuration at each doping level.
Because of Brillouin zone folding, the twists available are limited to the ones shown.
At $x=0.00$, the minimal gap is 2.6(1) eV, a little above the experimental gap of about 2.0 eV for the undoped cuprates.
The correction for calculating the gap at the $\Gamma$ point is around 0.5 eV\cite{wagner_effect_2014}, and so it is in good agreement with the experiment.
Meanwhile, at $x=0.25$, the gap clearly closes at $\left(\frac{1}{2},\frac{1}{2},0\right)$, which is in agreement with ARPES.

At $x=0.125$, it is possible that the gap closes near  the $\Gamma$ point $(0,0,0)$, which, accounting for band folding for the \cueight{} unit cell, is where the ARPES Fermi surface is be located. 
Because of the larger supercell, the stochastic errors could not be reduced below around 0.1 eV.
For \cco{}, it appears that the gap is not quite closed; however, systems close to metal-insulator transitions often suffer from larger finite size effects.
To check this, I also considered \ccoc{} at the $\Gamma$ point, which has a larger $c$-axis direction. 
It appears that the gap is either zero or near zero at the $\Gamma$ point for the \ccoc{} model, which increases the likelihood that the gap is actually closed at $x=0.125$.


There is also an excitation around 1 eV that appears upon doping.
This may correspond to new states seen in optical experiments at about that energy\cite{cilento_search_2013}.
So the spin polaron configuration has excitation properties largely in agreement with those seen in experiment for the doped cuprates.
Unfortunately the resolution is not high enough to comment on potential Fermi arcs.

\subsection{Polaron-phonon coupling}

While the focus of this article involves mainly the electronic degrees of freedom only, since the lowest energy state involves localized charge density, one might expect coupling to the lattice. 
Indeed, in the hybrid DFT calculations that also find a flipped ground state, the lattice reacts strongly to the presence of the hole, with relaxations of approximately 0.1 \AA.
In particular, the oxygen breathing mode is affected by the hole. 
This mode may be responsible for kinks in ARPES spectra\cite{garcia_through_2010}.
These effects warrant further investigation and will likely be important for a full description of the spin polaron.

\subsection{Mechanism for the spin polaron}

The behavior of the hole seen in the FN-DMC results warrants some explanation. 
Actually, the unusual behavior of magnetism in the cuprates begins even before the material is doped. 
Even in the undoped regime, there is a ferromagnetic-like interaction between the oxygen and copper. 
This is in contrast to the normal picture, even confirmed recently in FN-DMC calculations on VO$_2$\cite{zheng_computation_2015}, in which the interaction between the ligand and transition metal is antiferromagnetic-like. 
Once the material is doped, the copper spins change only by $\sim$0.1 Bohr magnetons, but the oxygen spin density is disrupted, as seen in Figure~\ref{fig:energetics}.

\begin{figure}
  \begin{tabular}{ccc}
    $\phi_\uparrow$ & $\phi_\downarrow$ & $\phi_\uparrow^2-\phi_\downarrow^2$ \\
    \includegraphics[width=0.25\columnwidth]{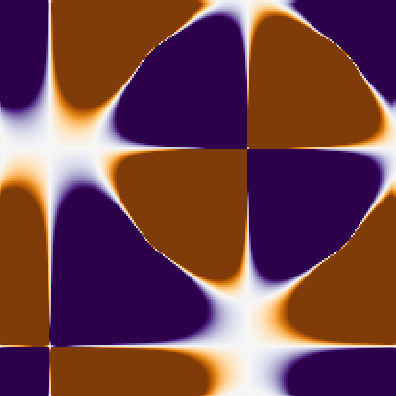} &
    \includegraphics[width=0.25\columnwidth]{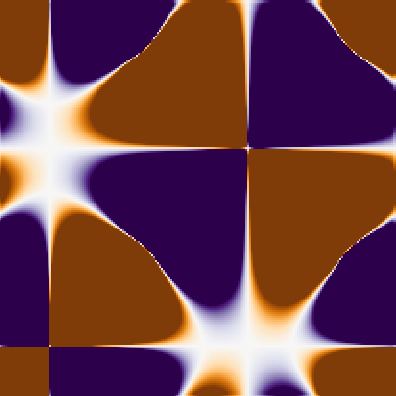} &
    \includegraphics[width=0.25\columnwidth]{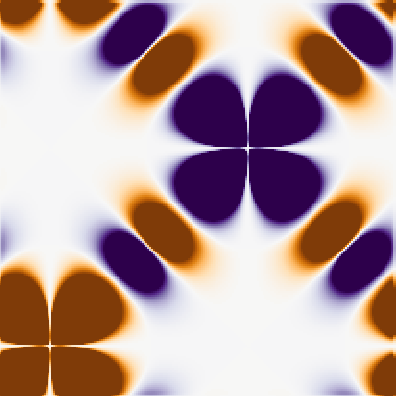} \\
    {\bf(a)}& {\bf(b)} & {\bf(c)} \\
    \includegraphics[width=0.25\columnwidth]{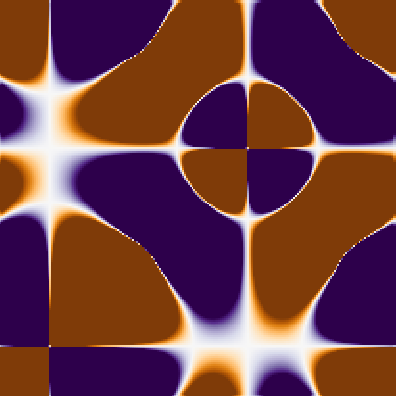} &
    \includegraphics[width=0.25\columnwidth]{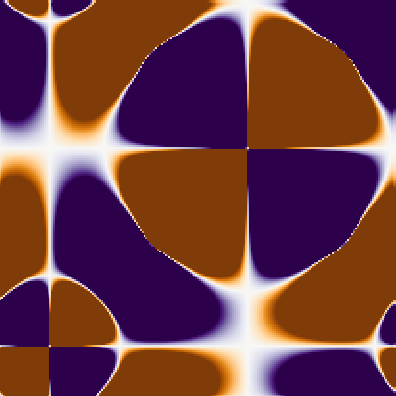} &
    \includegraphics[width=0.25\columnwidth]{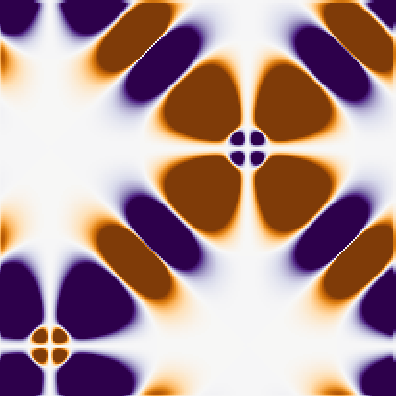} \\
    {\bf(d)} & {\bf(e)} & {\bf(f)} \\
  \end{tabular}
  \caption{Orbitals contributing to the spin density of a two formula unit cell of \cco{}.
    {\bf(a-c)} Lower energy orbital that contributes to most of the spin density on the copper atoms.
    {\bf(d-f)} Higher energy orbital that contributes to most of the spin density on the oxygen atoms.
    Blue is positive, red is negative.}
  \label{fig:orbitals}
\end{figure}

The unusual properties in the previous paragraph can be understood in terms of a pair of orbitals, diagrammed for a two copper unit cell in Figure~\ref{fig:orbitals}, with $x=0.00$ and in the AFM ordering.
The low-energy orbital is bonding-like between the oxygen and copper atoms, with more weight on one of the copper atoms due to the electron interaction. 
The second, higher energy, orbital is bonding for the spin minority copper atom and antibonding for the spin majority atom. 
When the spin density is calculated just from those orbitals, the low-energy orbital exhibits the antiferromagnetic relationship between the copper and oxygen spins.
However, the partial antibonding orbital has the opposite behavior and more spin density on the oxygen atoms. 
When the spin densities are summed, one obtains the ferromagnetic-like relationship between the oxygen and copper atoms.

On doping, the partially antibonding orbital is depopulated in PBE0. 
Since this orbital is responsible for the coupling between adjacent spins, the coupling between copper spins changes from antiferromagnetic to ferromagnetic. 
Electron correlations are important in determining that this orbital is depopulated upon introduction of a hole. 
In the FN-DMC calculations, we see that the spin up/spin down covariance on the copper atom around which the hole is centered decreases from -0.05 to -0.02.
This decrease is due to a decrease in the double occupancy of that copper atom. 
In addition, the orbital is partially antibonding, which costs kinetic energy.
The spin polaron, therefore, is stabilized by a balance between the kinetic energy and the interaction energy between electrons, which necessitates a correlated approach like FN-DMC.

\subsection{Charge and spin stripes}

\begin{figure}
  \begin{tabular}{cc}
  \includegraphics[width=0.45\columnwidth]{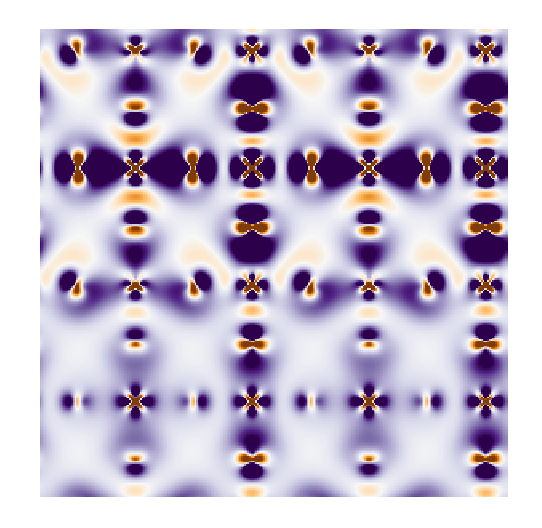} &
    \includegraphics[width=0.45\columnwidth]{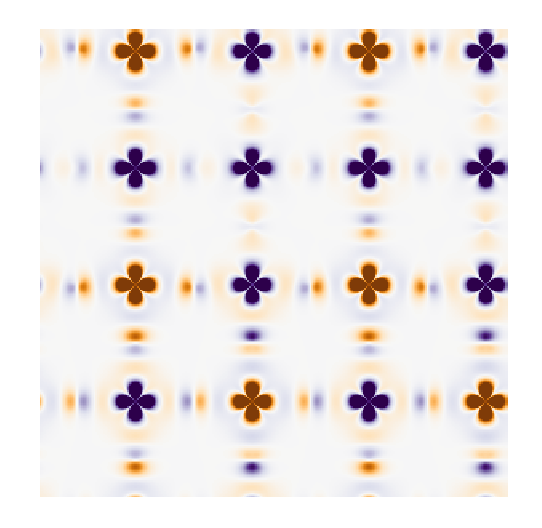} \\
    {\bf (a) } & {\bf (b) } \\
    \includegraphics[width=0.45\columnwidth]{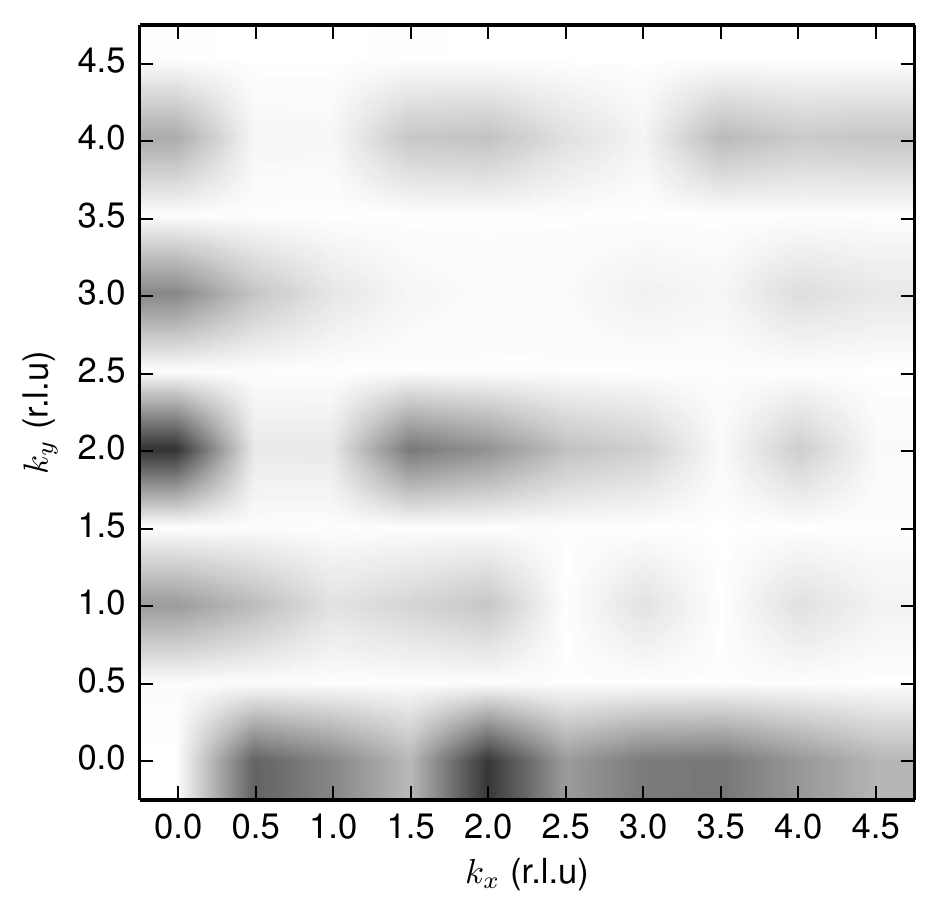} &
    \includegraphics[width=0.45\columnwidth]{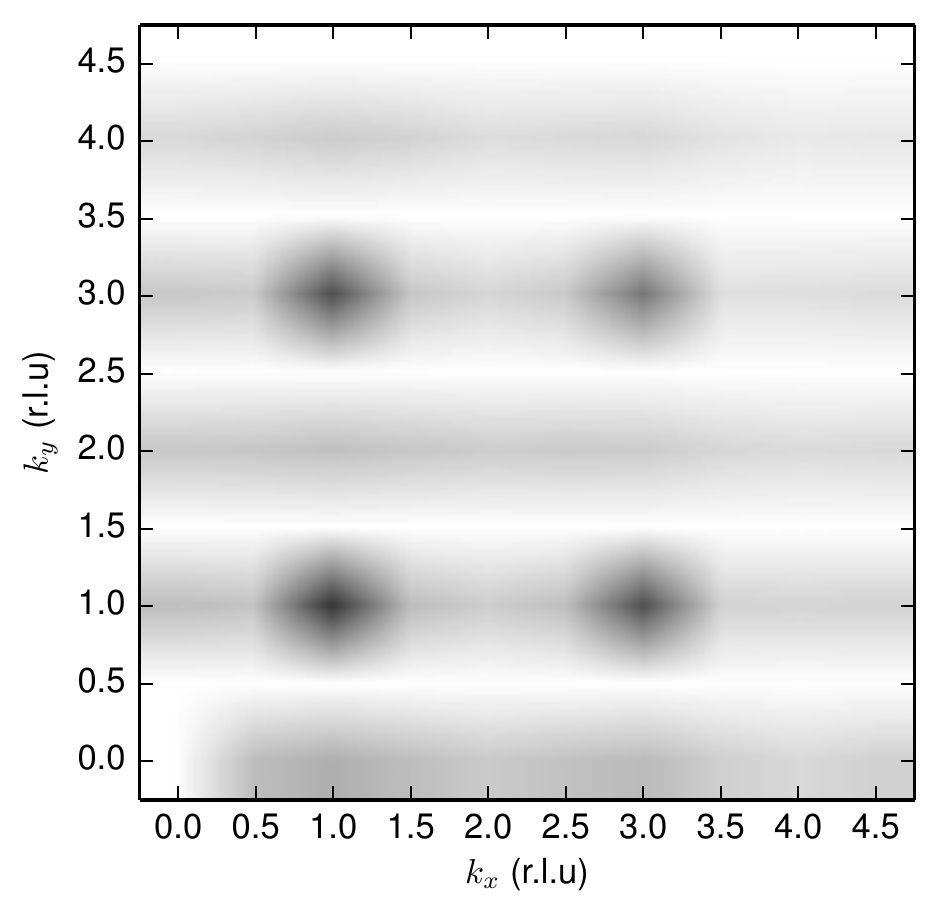} \\
    {\bf (c) } & {\bf (d) } \\
  \end{tabular}
  \caption{Charge {\bf(a)} and spin {\bf(b)} density in PBE0 for two holes in a 4$\times$4$\times$1 supercell. The corresponding Fourier transforms for charge {\bf(c)} and spin {\bf(d)} have a Bragg peak at (0,0.125) because of the periodicity in addition to the plotted values. The $k$ values are in units of the reciprocal lattice of the primitive tetragonal crystallographic cell.} 
  \label{fig:stripes}
\end{figure}

Patterson\cite{patterson_small_2008} noted that in hybrid DFT (B3LYP) calculations, it was possible to form charge and spin density waves using spin polarons.
In Figure~\ref{fig:stripes}, a similar structure is presented, along with the Fourier transform of the charge and spin.
Since the QMC calculations agree with DFT(PBE0) on the ground state density, it is likely good enough to analyze the properties of such a stripe system.
The stripe structure in Fig~\ref{fig:stripes} is in several ways quite close to that seen in neutron, x-ray, and STM experiments.
It matches the Bragg peak in the charge density at (0.25,0), as well as a d-wave intracell density, as seen in STM\cite{fujita_simultaneous_2014}.

If these stripes of spin polarons are the objects responsible for the stripes in the cuprates, then there should be a small peak in neutron diffraction at (1,0) and (0,1) due to the FM-like coupling between the copper atoms (Fig~\ref{fig:stripes}d), in addition to AFM-like (1,1) peaks.
This peak is a necessary prediction of this physics; if it is not present, then these objects cannot be responsible for the stripes.
On the other hand, the Fourier transform of the charge density should have small peaks at (0.5,0.25) from the periodicity of the charge density.
These small peaks are necessary for this structure to exist, and could be used to falsify these spin polarons as the origin of the stripes.

\section{Conclusion}

In summary, the microscopic physics of doped cuprates has been studied using state of the art quantum Monte Carlo techniques. 
The properties are very close to that seen in experiment, with the Fermi surface near that of the experiment, and optical excitations in agreement. 
The metal-insulator transition of the cuprates likely occurs between $x=0.00$ and $x=0.125$, which means that this transition has been captured accurately within a first-principles method, a new result.
In a few years, it will likely be possible to use these techniques to study the metal-insulator transition in detail.
Much of the phenomenology of the cuprates is consistent with the picture emerging from the FN-DMC results.
This is a particularly exciting result: since the FN-DMC methodology employed here does not use any effective parameterization of the interactions, it is a predictive methodology that could be helpful in the search for materials with similar physics.

The picture emerging from the FN-DMC calculations is that of a spin polaron. 
The particular polaron found in this work was to my knowledge first proposed by Emery and Reiter\cite{emery_mechanism_1988} early on, and has been considered by a number of authors since then\cite{patterson_small_2008}.
The uniqueness of this study is that 1) it is truly first principles and explicitly correlated, with no adjustable parameters, 2) it was shown that the spin polaron can lead to a similar Fermi surface structure as seen in experiment, and 3) there are new predictions which allow the proposal to be evaluated experimentally.
This spin polaron can be further studied using quantum Monte Carlo techniques, and it will be fruitful to study the properties of this quasiparticle; for example, the interaction between two spin polarons.
This spin polaron is not stable in the closely related material \cnoc{}, and thus may be one reason for the uniqueness of the cuprates.

I would like to thank the many people who contributed comments to this paper, in particular
Hitesh Changlani,
David Ceperley,
J.C. Seamus Davis,
Awadhesh Narayan, Mike Norman,
and
Huihuo Zheng.
This material is based upon work supported by the U.S. Department of Energy, Office of Science, Office of Advanced Scientific Computing Research, Scientific Discovery through Advanced Computing (SciDAC) program under Award Number FG02-12ER46875.
Computational resources were provided by the DOE INCITE SuperMatSim and PhotoSuper programs.

\bibliographystyle{unsrt}
\bibliography{cuprate_doping}

\begin{thebibliography}{10}

\bibitem{zhang_effective_1988}
F.~C. Zhang and T.~M. Rice.
\newblock Effective {Hamiltonian} for the superconducting {Cu} oxides.
\newblock {\em Physical Review B}, 37(7):3759--3761, March 1988.

\bibitem{emery_mechanism_1988}
V.~J. Emery and G.~Reiter.
\newblock Mechanism for high-temperature superconductivity.
\newblock {\em Physical Review B}, 38(7):4547--4556, September 1988.

\bibitem{martin_electronic_1996}
Richard~L. Martin.
\newblock Electronic localization in the cuprates.
\newblock {\em Physical Review B}, 53(23):15501--15512, June 1996.

\bibitem{mott_spin-polaron_1990}
N.F. Mott.
\newblock The spin-polaron theory of high-{T} c superconductivity.
\newblock {\em Advances in Physics}, 39(1):55--81, February 1990.

\bibitem{lau_high-spin_2011}
Bayo Lau, Mona Berciu, and George~A. Sawatzky.
\newblock High-{Spin} {Polaron} in {Lightly} {Doped} {CuO}2 {Planes}.
\newblock {\em Physical Review Letters}, 106(3):036401, January 2011.

\bibitem{patterson_small_2008}
C.~H. Patterson.
\newblock Small polarons and magnetic antiphase boundaries in
  {Ca}(2-x){NaxCuO}2cl2 (x=0.06,0.12): {Origin} of striped phases in cuprates.
\newblock {\em Physical Review B}, 77(9):094523, March 2008.

\bibitem{hozoi_renormalization_2007}
L.~Hozoi, S.~Nishimoto, and C.~de~Graaf.
\newblock Renormalization of quasiparticle hopping integrals by spin
  interactions in layered copper oxides.
\newblock {\em Physical Review B}, 75(17):174505, May 2007.

\bibitem{hashimoto_energy_2014}
Makoto Hashimoto, Inna~M. Vishik, Rui-Hua He, Thomas~P. Devereaux, and Zhi-Xun
  Shen.
\newblock Energy gaps in high-transition-temperature cuprate superconductors.
\newblock {\em Nature Physics}, 10(7):483--495, July 2014.

\bibitem{dagotto_correlated_1994}
Elbio Dagotto.
\newblock Correlated electrons in high-temperature superconductors.
\newblock {\em Reviews of Modern Physics}, 66(3):763--840, July 1994.

\bibitem{damascelli_angle-resolved_2003}
Andrea Damascelli, Zahid Hussain, and Zhi-Xun Shen.
\newblock Angle-resolved photoemission studies of the cuprate superconductors.
\newblock {\em Reviews of Modern Physics}, 75(2):473--541, April 2003.

\bibitem{kresin_colloquium:_2009}
V.~Z. Kresin and S.~A. Wolf.
\newblock Colloquium: {Electron}-lattice interaction and its impact on high
  {T}\_c superconductivity.
\newblock {\em Reviews of Modern Physics}, 81(2):481--501, April 2009.

\bibitem{scalapino_common_2012}
D.~J. Scalapino.
\newblock A common thread: {The} pairing interaction for unconventional
  superconductors.
\newblock {\em Reviews of Modern Physics}, 84(4):1383--1417, October 2012.

\bibitem{tsuei_pairing_2000}
C.~C. Tsuei and J.~R. Kirtley.
\newblock Pairing symmetry in cuprate superconductors.
\newblock {\em Reviews of Modern Physics}, 72(4):969--1016, October 2000.

\bibitem{razzoli_fermi_2010}
E~Razzoli, Y~Sassa, G~Drachuck, M~Månsson, A~Keren, M~Shay, M~H Berntsen,
  O~Tjernberg, M~Radovic, J~Chang, S~Pailhès, N~Momono, M~Oda, M~Ido, O~J
  Lipscombe, S~M Hayden, L~Patthey, J~Mesot, and M~Shi.
\newblock The {Fermi} surface and band folding in {La}(2-x){SrxCuO}4, probed by
  angle-resolved photoemission.
\newblock {\em New Journal of Physics}, 12(12):125003, December 2010.

\bibitem{doiron-leyraud_quantum_2007}
Nicolas Doiron-Leyraud, Cyril Proust, David LeBoeuf, Julien Levallois,
  Jean-Baptiste Bonnemaison, Ruixing Liang, D.~A. Bonn, W.~N. Hardy, and Louis
  Taillefer.
\newblock Quantum oscillations and the {Fermi} surface in an underdoped
  high-{Tc} superconductor.
\newblock {\em Nature}, 447(7144):565--568, May 2007.

\bibitem{cilento_search_2013}
F~Cilento, S~Dal Conte, G~Coslovich, F~Banfi, G~Ferrini, H~Eisaki, M~Greven,
  A~Damascelli, D~van~der Marel, F~Parmigiani, and C~Giannetti.
\newblock In search for the pairing glue in cuprates by non-equilibrium optical
  spectroscopy.
\newblock {\em Journal of Physics: Conference Series}, 449:012003, July 2013.

\bibitem{fujita_simultaneous_2014}
K.~Fujita, Chung~Koo Kim, Inhee Lee, Jinho Lee, M.~H. Hamidian, I.~A. Firmo,
  S.~Mukhopadhyay, H.~Eisaki, S.~Uchida, M.~J. Lawler, E.-A. Kim, and J.~C.
  Davis.
\newblock Simultaneous {Transitions} in {Cuprate} {Momentum}-{Space} {Topology}
  and {Electronic} {Symmetry} {Breaking}.
\newblock {\em Science}, 344(6184):612--616, May 2014.

\bibitem{mashima_electronic_2006}
H.~Mashima, N.~Fukuo, Y.~Matsumoto, G.~Kinoda, T.~Kondo, H.~Ikuta, T.~Hitosugi,
  and T.~Hasegawa.
\newblock Electronic inhomogeneity of heavily overdoped {Bi}(2-x){PbxSr}2cuoy
  studied by low-temperature scanning tunneling microscopy/spectroscopy.
\newblock {\em Physical Review B}, 73(6):060502, February 2006.

\bibitem{wise_charge-density-wave_2008}
W.~D. Wise, M.~C. Boyer, Kamalesh Chatterjee, Takeshi Kondo, T.~Takeuchi,
  H.~Ikuta, Yayu Wang, and E.~W. Hudson.
\newblock Charge-density-wave origin of cuprate checkerboard visualized by
  scanning tunnelling microscopy.
\newblock {\em Nature Physics}, 4(9):696--699, September 2008.

\bibitem{tranquada_evidence_1995}
J.~M. Tranquada, B.~J. Sternlieb, J.~D. Axe, Y.~Nakamura, and S.~Uchida.
\newblock Evidence for stripe correlations of spins and holes in copper oxide
  superconductors.
\newblock {\em Nature}, 375(6532):561--563, June 1995.

\bibitem{abbamonte_spatially_2005}
P.~Abbamonte, A.~Rusydi, S.~Smadici, G.~D. Gu, G.~A. Sawatzky, and D.~L. Feng.
\newblock Spatially modulated '{Mottness}' in {La}2-{xBaxCuO}4.
\newblock {\em Nature Physics}, 1(3):155--158, December 2005.

\bibitem{rivero_general_2013}
Pablo Rivero, Ibério de~P.~R.Moreira, Ricardo Grau-Crespo, Sambhu~N. Datta,
  and Francesc Illas.
\newblock General model for explicitly hole-doped superconductor parent
  compounds: {Electronic} structure of {Ca}2-{xNaxCuO}2cl2 as a case study.
\newblock {\em Physical Review B}, 88(8):085108, August 2013.

\bibitem{pesant_dft+u_2011}
Simon Pesant and Michel Cote.
\newblock {DFT}+{U} study of magnetic order in doped {La}2cuo4 crystals.
\newblock {\em Physical Review B}, 84(8):085104, August 2011.

\bibitem{weber_strength_2010}
Cedric Weber, Kristjan Haule, and Gabriel Kotliar.
\newblock Strength of correlations in electron- and hole-doped cuprates.
\newblock {\em Nature Physics}, 6(8):574--578, August 2010.

\bibitem{foyevtsova_ab_2014}
Kateryna Foyevtsova, Jaron~T. Krogel, Jeongnim Kim, P. R. C. Kent, Elbio
  Dagotto, and Fernando~A. Reboredo.
\newblock Ab initio {Quantum} {Monte} {Carlo} {Calculations} of {Spin}
  {Superexchange} in {Cuprates}: {The} {Benchmarking} {Case} of {Ca}2cuo3.
\newblock {\em Physical Review X}, 4(3):031003, July 2014.

\bibitem{wagner_effect_2014}
Lucas~K. Wagner and Peter Abbamonte.
\newblock Effect of electron correlation on the electronic structure and
  spin-lattice coupling of high-{Tc} cuprates: {Quantum} {Monte} {Carlo}
  calculations.
\newblock {\em Physical Review B}, 90(12):125129, September 2014.

\bibitem{burkatzki_energy-consistent_2007}
M.~Burkatzki, C.~Filippi, and M.~Dolg.
\newblock Energy-consistent pseudopotentials for quantum {Monte} {Carlo}
  calculations.
\newblock {\em The Journal of Chemical Physics}, 126(23):234105--234105--8,
  June 2007.

\bibitem{burkatzki_energy-consistent_2008}
M.~Burkatzki, Claudia Filippi, and M.~Dolg.
\newblock Energy-consistent small-core pseudopotentials for 3d-transition
  metals adapted to quantum {Monte} {Carlo} calculations.
\newblock {\em The Journal of Chemical Physics}, 129(16):164115--164115--7,
  October 2008.

\bibitem{dovesi_crystal:_2005}
Roberto Dovesi, Roberto Orlando, Bartolomeo Civalleri, Carla Roetti, Victor~R.
  Saunders, and Claudio~M. Zicovich-Wilson.
\newblock {CRYSTAL}: a computational tool for the ab initio study of the
  electronic properties of crystals.
\newblock {\em Zeitschrift für Kristallographie}, 220(5-6-2005):571--573, May
  2005.

\bibitem{wagner_qwalk:_2009}
Lucas~K. Wagner, Michal Bajdich, and Lubos Mitas.
\newblock {QWalk}: {A} quantum {Monte} {Carlo} program for electronic
  structure.
\newblock {\em Journal of Computational Physics}, 228(9):3390--3404, May 2009.

\bibitem{perdew_generalized_1996}
John~P. Perdew, Kieron Burke, and Matthias Ernzerhof.
\newblock Generalized {Gradient} {Approximation} {Made} {Simple}.
\newblock {\em Physical Review Letters}, 77(18):3865, October 1996.

\bibitem{adamo_toward_1999}
Carlo Adamo and Vincenzo Barone.
\newblock Toward reliable density functional methods without adjustable
  parameters: {The} {PBE}0 model.
\newblock {\em The Journal of Chemical Physics}, 110(13):6158, 1999.

\bibitem{casula_beyond_2006}
Michele Casula.
\newblock Beyond the locality approximation in the standard diffusion {Monte}
  {Carlo} method.
\newblock {\em Physical Review B}, 74(16), October 2006.

\bibitem{garcia_through_2010}
D.~R. Garcia and A.~Lanzara.
\newblock Through a {Lattice} {Darkly}: {Shedding} {Light} on
  {Electron}-{Phonon} {Coupling} in the {High} {Tc} {Cuprates}.
\newblock {\em Advances in Condensed Matter Physics}, 2010:1--23, 2010.

\bibitem{zheng_computation_2015}
Huihuo Zheng and Lucas~K. Wagner.
\newblock Computation of the {Correlated} {Metal}-{Insulator} {Transition} in
  {Vanadium} {Dioxide} from {First} {Principles}.
\newblock {\em Physical Review Letters}, 114(17):176401, April 2015.

\end{thebibliography}

\end{document}